\documentclass[twocolumn,english,aps,prl,reprint,nofootinbib,superscriptaddress]{revtex4-2}
\pdfoutput=1
\usepackage{svg}
\usepackage[T1]{fontenc}
\usepackage[latin9]{inputenc}
\usepackage{amsmath}
\usepackage{amssymb}
\usepackage{graphicx}
\usepackage{enumerate}
\usepackage{braket}
\usepackage{color}
\usepackage{float}

\makeatletter
\@ifundefined{textcolor}{}
{%
 \definecolor{BLACK}{gray}{0}
 \definecolor{WHITE}{gray}{1}
 \definecolor{RED}{rgb}{1,0,0}
 \definecolor{GREEN}{rgb}{0,1,0}
 \definecolor{BLUE}{rgb}{0,0,1}
 \definecolor{CYAN}{cmyk}{1,0,0,0}
 \definecolor{MAGENTA}{cmyk}{0,1,0,0}
 \definecolor{YELLOW}{cmyk}{0,0,1,0}
 \definecolor{PURPLE}{rgb}{0.7,0,0.7}
 \definecolor{dgreen}{rgb}{0,0.6,0}
}

\newcommand{\tm}[1]{\mathcal{#1}}

\usepackage{pslatex}

\usepackage{babel}

\usepackage{hyperref}

\hypersetup{
    colorlinks,
    linkcolor={black},
    citecolor={black},
    urlcolor={blue},
}

\makeatother

\begin{document}

\title{Measurement of the H$^3\Delta_1$ Radiative Lifetime in ThO}
\author{D. G. Ang}
\email[Corresponding author, ]{danielang@g.harvard.edu}
\affiliation{Department of Physics, Harvard University, Cambridge, Massachusetts 02138, USA}
\author{C. Meisenhelder}
\affiliation{Department of Physics, Harvard University, Cambridge, Massachusetts 02138, USA}
\author{C. D. Panda}
\altaffiliation[Present address: ]{Department of Physics, University of California, Berkeley, Berkeley, California 94720, USA.}
\affiliation{Department of Physics, Harvard University, Cambridge, Massachusetts 02138, USA}
\author{X. Wu}
\affiliation{Department of Physics, Harvard University, Cambridge, Massachusetts 02138, USA}
\affiliation{Department of Physics, University of Chicago, Chicago, Illinois 60637, USA}
\author{D. DeMille}
\affiliation{Department of Physics, University of Chicago, Chicago, Illinois 60637, USA}
\author{J. M. Doyle}
\affiliation{Department of Physics, Harvard University, Cambridge, Massachusetts 02138, USA}
\author{G. Gabrielse}
\email[Corresponding author, ]{ gerald.gabrielse@northwestern.edu}
\affiliation{Center for Fundamental Physics, Northwestern University, Evanston, Illinois 60208, USA}
\date{\today}

\begin{abstract}
The best limit on the electron electric dipole moment (eEDM) comes from the ACME II experiment (Nature \textbf{562} (2018), 355-360) which probes physics beyond the Standard Model at energy scales well above 1 TeV. ACME II measured the eEDM by monitoring electron spin precession in a cold beam of the metastable H$^3\Delta_1$ state of thorium monoxide (ThO) molecules, with an observation time $\tau \approx 1$ ms for each molecule. We report here a new measurement of the lifetime of the ThO (H$^3\Delta_1$) state,  $\tau_H = 4.2\pm 0.5$ ms. Using an apparatus within which $\tau \approx \tau_H$ will enable a substantial reduction in uncertainty of an eEDM measurement.
\end{abstract}

\maketitle

\section{Introduction}

In 2018, the ACME II measurement set the current best upper limit on the electron electric dipole moment (EDM), $|d_e|<1.1\times10^{-29}\,$e$\cdot$cm~\cite{ACMECollaboration2018}, an order of magnitude improvement over previous measurements~\cite{Baron2014,Cairncross2017}. This measurement set stringent constraints on various scenarios of CP-violating new physics in the 3-30 TeV mass range~\cite{Cesarotti2019,Panico2019}. ACME III, a new generation of the ACME experiment, is now being launched with the aim of measuring the electron EDM at higher precision and thus probing for new physics at even higher energy scales. 

ACME measured the electron EDM by performing a spin precession measurement in a beam of thorium monoxide molecules (ThO) in the metastable H($^3\Delta_1)$ electronic state. This state was selected for having a large effective internal electric field ($\mathcal{E}_{\mathrm{eff}} \sim$78 GV/cm~\cite{Denis2016,Skripnikov2016}) that greatly amplifies the interaction of the laboratory electric field with the electron EDM, $d_e$, its unusually small magnetic moment, and its $\Omega$-doublet energy level structure which provides a powerful means to diagnose and reject systematic errors~\cite{Kirilov2013}. The statistical sensitivity of the measurement (given that the shot-noise limited sensitivity is attained~\cite{Panda2019}) is
\begin{equation}
    \delta d_e = 
    \frac{1}{2 \tau \mathcal{E}_{\mathrm{eff}} \sqrt{N}},
    \label{eqn:edmunc}
\end{equation}
where $\tau$ is the coherence time, $\mathcal{E}_{\mathrm{eff}}$ is the internal effective electric field, and $N$ is the number of molecules detected in the measurement. Increasing $\tau$ or $\sqrt{N}$ could produce better sensitivity, the former being especially attractive because there is no square root involved.

The radiative lifetime of the $H$-state  ($\tau_H$) is important because it limits the coherence time ($\tau$) that can be achieved. Attempts to measure this lifetime in a gas cell established that it was long enough for a practical choice of a $\tau\approx 1$ ms coherence time for the ACME I and ACME II measurements, but also that collisions and observed multi-exponential decays complicated a reliable determination of the lifetime~\cite{Vutha2010,vuthathesis,yatthesis}. To optimally design a new generation measurement, the molecular beam measurement reported here was carried out. Collisions are not a factor in the low density beam, and the radiation decay over time could be directly studied. ThO molecules are excited into the $H$-state using laser light at various positions along a molecular beam.  The population remaining after radiative decay to the ground state is then probed at a fixed detector position. A much longer $\tau_H$ is deduced than was previously estimated.   

The method and observations are discussed in Sec.~\ref{sec:expmethod}, the uncertainties in Sec.~\ref{sec:systematics}, and the results in Sec.~\ref{sec:results}.  Implications for an improved electric dipole moment are presented in Sec.~\ref{sec:implications}.

\section{Method and observations}
\label{sec:expmethod}

\begin{figure*}[htp]
\centerline{
\includegraphics[width = 7in]{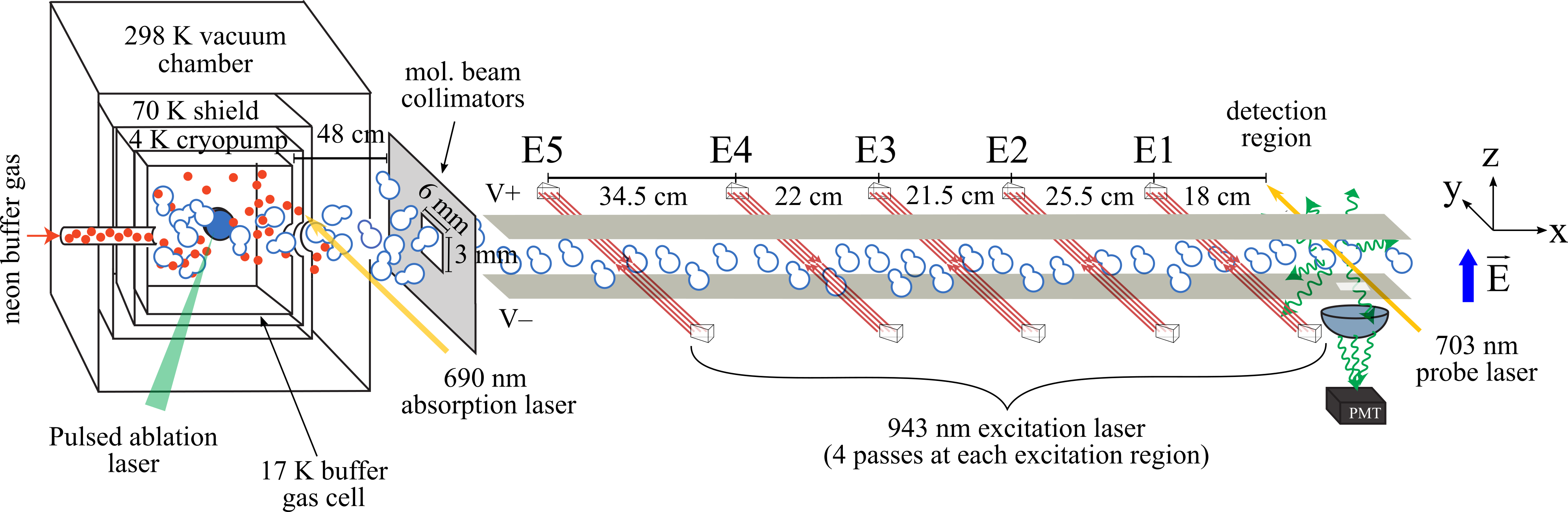}
}\caption{Experimental setup (not to scale) used to probe the $H$ state lifetime.}\label{fig:exp_diag}
\end{figure*}

Laser ablation and a cryogenic buffer gas produce pulses of ThO molecules with a mean longitudinal velocity of $\sim$210 m/s \cite{Hutzler2011a} (Fig. \ref{fig:exp_diag}). 
The molecules are mostly in their ground electronic ($X$) and lowest vibrational states, with a rotational temperature of $\sim$4 K. A laser tuned to the 690 nm  $X-C$ electronic transition produces an optical absorption signal just after the buffer gas cell which is used as part of the determination of the molecular velocity.  About 48 cm downstream from the cell aperture (which has a diameter of 5 mm), 6 mm horizontal and 3 mm vertical collimators control the size and distribution of transverse velocities in the molecular beam, giving a $1\sigma$ Doppler width of $6 \pm 1$ MHz along the $y$-axis for a laser wavelength of 943 nm. The molecular pulses are $\sim$0.5 ms in duration as they leave the source but expand to a bit more than 2 ms by the time they arrive at a detection region that is 1.78 m away.  

\begin{figure}[htpb]
\begin{center}
\includegraphics[width=3.25in]{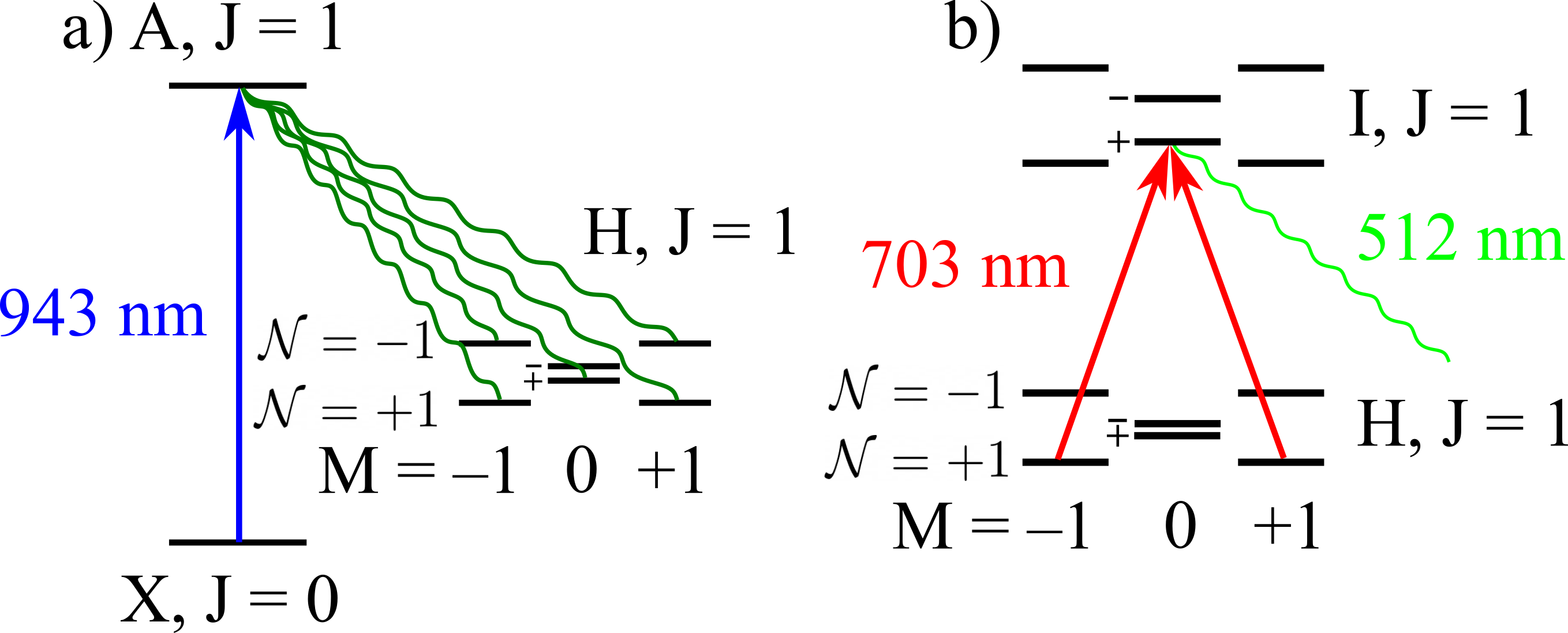}
\caption{a) Optical pumping excitation of the $H$-state in an electric field. b) Detection of fluorescence after a second excitation.}
\label{fig:opticalpumping}
\end{center}
\end{figure}

A pair of $\sim$1.3 m long parallel plates produces a $\sim$38 V/cm electric field that is vertical along $\hat{z}$, perpendicular to the molecular beam direction $\hat{x}$.  
Within this field, at one of 5 nominally-identical excitation regions (labeled E1-E5 in Fig.~\ref{fig:exp_diag}), an excitation laser transfers molecules into the metastable $H$ state which is fully polarized within the applied electric field~\cite{Vutha2010}. The five excitation regions are located at distances from the detection region $L_i$, measured from the center of the beam volume occupied by multiple passes of the excitation lasers (Table \ref{tab:pp}).
The 943 nm excitation laser is linearly polarized along $\hat{z}$ and excites the transition $\ket{X, J=0}\rightarrow \ket{A, J=1}$. About 30\% of the molecules spontaneously decay to an incoherent mixture of $M$ and $\tm{N}$ states within the $\ket{H, J=1}$ manifold (Fig.~\ref{fig:opticalpumping}a)~\cite{spaun}. Here, $J$ is the angular momentum quantum number and $M$ is its projection along the quantization axis $\hat{z}$. $\tm{N}=\pm 1$ correspond to states of opposite orientation of $\mathcal{E}_{\mathrm{eff}}$ with respect to the applied laboratory electric field. All used states have vibrational quantum number $\nu = 0$. 

The molecules freely propagate down the beam line while undergoing radiative decay from the metastable $H$ state to the stable ground state $X$.  The number of $H$ state molecules reaching the detection region decreases exponentially as $e^{-t/\tau_H}$ with lifetime $\tau_H$, where $t$ is the time between excitation and detection.  This remaining population is probed by optically pumping the $\ket{H, J=1}\rightarrow \ket{I, J=1}$ transition using a 703 nm probe laser, linearly polarized along $\hat{x}$ (Fig.~\ref{fig:opticalpumping}b). The $I$-state is short-lived and rapidly decays back to the ground state, producing 512 nm photons which are detected using a photomultiplier. The intensities are normalized to that observed for excitation at E1 and fit to an exponential decay curve  to obtain $\tau_H$.

\begin{table}[htbp]
\centering
\begin{tabular}{|c|c|}
\hline
\textbf{No.} & \textbf{\begin{tabular}[c]{@{}c@{}}Distance to detection \\ region $L_i$ (cm)\end{tabular}} \\ \hline
E1          & $18.0 \pm 1.0$                                                                                         \\ \hline
E2          & $43.5 \pm 1.0 $                                                                                        \\ \hline
E3          & $65.0 \pm 1.0 $                                                                                          \\ \hline
E4          & $87.0  \pm 1.0 $                                                                                         \\ \hline
E5          & $121.5 \pm 1.0 $                                                                                      \\ \hline
\end{tabular}
\caption{Distances between the excitation regions (E\#) and the detection region.}
\label{tab:pp}
\end{table}

The excitation laser (Toptica DL Pro with Roithner RLT0940-300GS diode followed by a Toptica BoosTA tapered amplifier) is locked to an iodine-clock-stabilized laser using a slow scanning cavity transfer lock, resulting in a laser linewidth (FWHM) of $\sim$2 MHz. Software-controlled acousto-optic modulators (AOMs) coupled to optical fibers enable rapid switching of the optical path of the laser between different Es. Each excitation region contains 65 mW ($\pm 10\%$) of laser power that is quadruple-passed through the molecular beam using a pair of prisms to improve saturation of the optical pumping, resulting in $\sim$230 mW of total circulating power (after accounting for transmission losses from the vacuum windows). The data was acquired in two data sets. For the second data set, replacement of an optical isolator for the excitation laser and aging of the TA resulted in a loss of $\sim$20\% laser power compared to the first data set. The laser is linearly polarized along the $z$-axis and propagates along the $y$-axis (see Fig.~\ref{fig:exp_diag}). Each excitation region also contains an independent set of optics to expand the laser beam to a $1/e^2$ height and width of $\sim$1 cm and $\sim$0.1 cm respectively. This height was selected based on the measured height of the molecular beam at each excitation region of $\lesssim 0.8$ cm. Ensuring that we address all of the molecules reduces the complicating effects of molecular beam divergence.

The probe laser (M Squared SolsTiS TiSapph) is locked to a second, stabilized low-powered diode laser using a delay line transfer lock scheme~\cite{delaylinelock}\cite{pandathesis}. The second laser is stabilized to a high-finesse optical cavity using a Pound-Drever-Hall locking scheme~\cite{Drever}. This results in the probe laser having a linewidth of $\sim$20 kHz. The first data set was taken with a laser power of 160 mW, and the second data set with 240 mW. At the detection region, the probe laser beam is expanded to the same size as the excitation laser and linearly polarized along the $x$-axis.

\begin{figure}[htpb!]
\begin{center}
\includegraphics[width=3.25in]{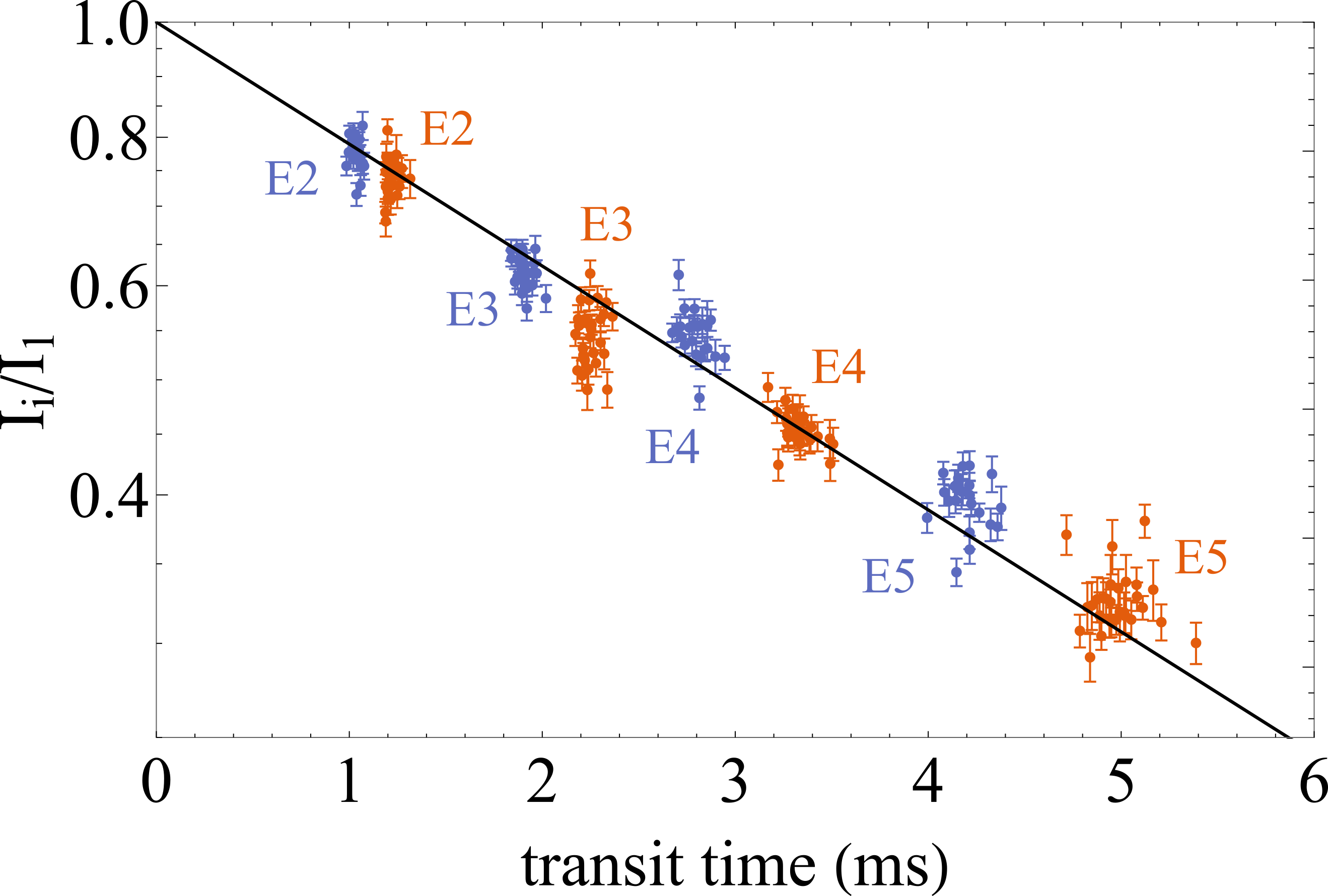}
\vspace{0mm}
\caption{Semilog plot of the intensity ratios as a function of transit time between the excitation regions and fluorescence  detector, with a fit to Eq.~\ref{eq:fit}. The colors indicate two data sets taken at widely separated times, before (blue) and after (orange) an ablation target change that resulted in a change in ThO velocity.}
\label{fig:twodatafit}
\end{center}
\end{figure}

The molecular pulse intensity typically varies by 10-15\% from pulse to pulse. This signal also slowly decays, dropping by as much as 50\% after 5-15 minutes. We revive the signal by moving the ablation laser to a different location on the ThO\textsubscript{2} target~\cite{Baron2017}. 
To suppress the effect of these changes, we use the large signal from excitation at E1 for normalization. In the first data set, fluorescence from 64 consecutive molecular pulses were averaged into traces of length 16 ms. (In the second data set, 32 pulses were averaged to enable quicker change and optimization of the laser ablation spots on the ThO\textsubscript{2} target.) For each trace, we subtract a background by sampling the first 3 ms and last 4 ms and integrate the molecular pulse signal by sampling the central 9 ms. We switch between acquiring E1 data and one of E2-5 every 7-9 seconds, resulting in two groups of several traces, from which we calculate the relative intensity $I_i/I_1$. Figure \ref{fig:twodatafit} is a semilog plot of the observed relative intensities at the excitation regions $i = 2, ..., 5$. The ablation target used to produce the ThO molecules was changed between the first (blue points) and second (orange points) datasets. The change in velocity for a new target is a familiar consequence \cite{Hutzler2011a}. 

The measured relative intensities are fit to 
\begin{equation}
    \frac{I_i}{I_1} =  \, \exp{\left(
   -\frac{L_i-L_1}{v \, \tau_H}
    \right)}
    \label{eq:fit}
\end{equation}
to obtain the best fit straight line shown on the semilog plot. The measured $L_i$ are in Table \ref{tab:pp}.  The molecular beam velocity $v$ is measured by subtracting the arrival times of the center of mass of the fluorescence trace at the detection region and the optical absorption trace from the 690 nm laser placed just after the ablation cell (as shown in Fig.~\ref{fig:comfl}). The statistical uncertainty in $\tau_H$ from the fit (0.02 ms) is computed using standard error estimation procedures from maximum likelihood estimation~\cite{lyons1989statistics}. A computation using bootstrapped datasets~\cite{Efron1986} yields the same value. As is obvious from Fig.~\ref{fig:twodatafit}, and a reduced chi-squared value of of 4.4, the statistical uncertainty is small compared to other sources of uncertainty that are discussed in the next section. 

\section{Systematic Uncertainties}
\label{sec:systematics}

The largest of the systematic uncertainties comes from possible variations in excitation efficiencies at the excitation regions, given that Eq.~\ref{eq:fit} assumes an equal number of $H$-state molecules leave each excitation region. In practice, there could be variations between the excitation regions due to velocity dispersion, differences in laser illumination and saturation, or misalignments.  
Fluctuations from one excitation region to the next can be represented as a set of excitation efficiencies $\eta_i$ that are slightly different for each excitation region.  

To estimate the spread of excitation efficiencies,
the detected signal as a function of laser power $P$ at excitation point $i$ is fit to $S = S_{max} [1-\exp(-P/P_s)]$ to extract the saturation power $P_s$ and the extrapolated maximum signal $S_{max}$.  The efficiency is $\eta_i= S/S_{max}$ for the laser power $P$ used for the lifetime measurement. We obtain a spread in excitation efficiencies of up to 7\% from the measured values and their uncertainties.

To determine the effect of this variation in excitation efficiency on the uncertainty of $\tau_H$, the data is fit for each of a large set of efficiencies selected randomly from a Gaussian distribution with a 7\% standard deviation.
The distribution of the resulting $\tau_H$ values has a standard deviation of 0.4 ms.  This value is used as the estimated systematic uncertainty from differing excitation efficiencies at the excitation regions.

A smaller contribution that only slightly increases the lifetime uncertainty comes from uncertainty in the background to be subtracted before the signal is fit to an exponential decay. Detuning the excitation laser far from resonance established that the background is below 3\% of $I_1$, which corresponds to a 0.2 ms uncertainty.   

To get substantial laser saturation, the laser beams cross the molecular beam four times at each excitation region. The resulting width of each interaction region makes it possible to localize the distance between the excitation regions and the detection region to about 1 cm. To estimate the resulting uncertainty in $\tau_H$, the data is fit to fitting functions like Eq.~\ref{eq:fit} but with $L_i-L_1$ in each case offset by a value from a Gaussian distribution that is $\pm \sqrt{2}$ cm wide on average.  From this we learn that the 1 cm uncertainty most likely contributes an uncertainty of $\pm 0.04$ ms in uncertainty to $\tau_H$, which can be neglected compared to the other uncertainties. The key here is that the 1 cm uncertainty is fractionally small compared to the longer $L_i-L_1$ values corresponding to excitation regions which are given more weight in the fit.     

\begin{figure}[htpb]
\begin{center}
\includegraphics[width=3.25in]{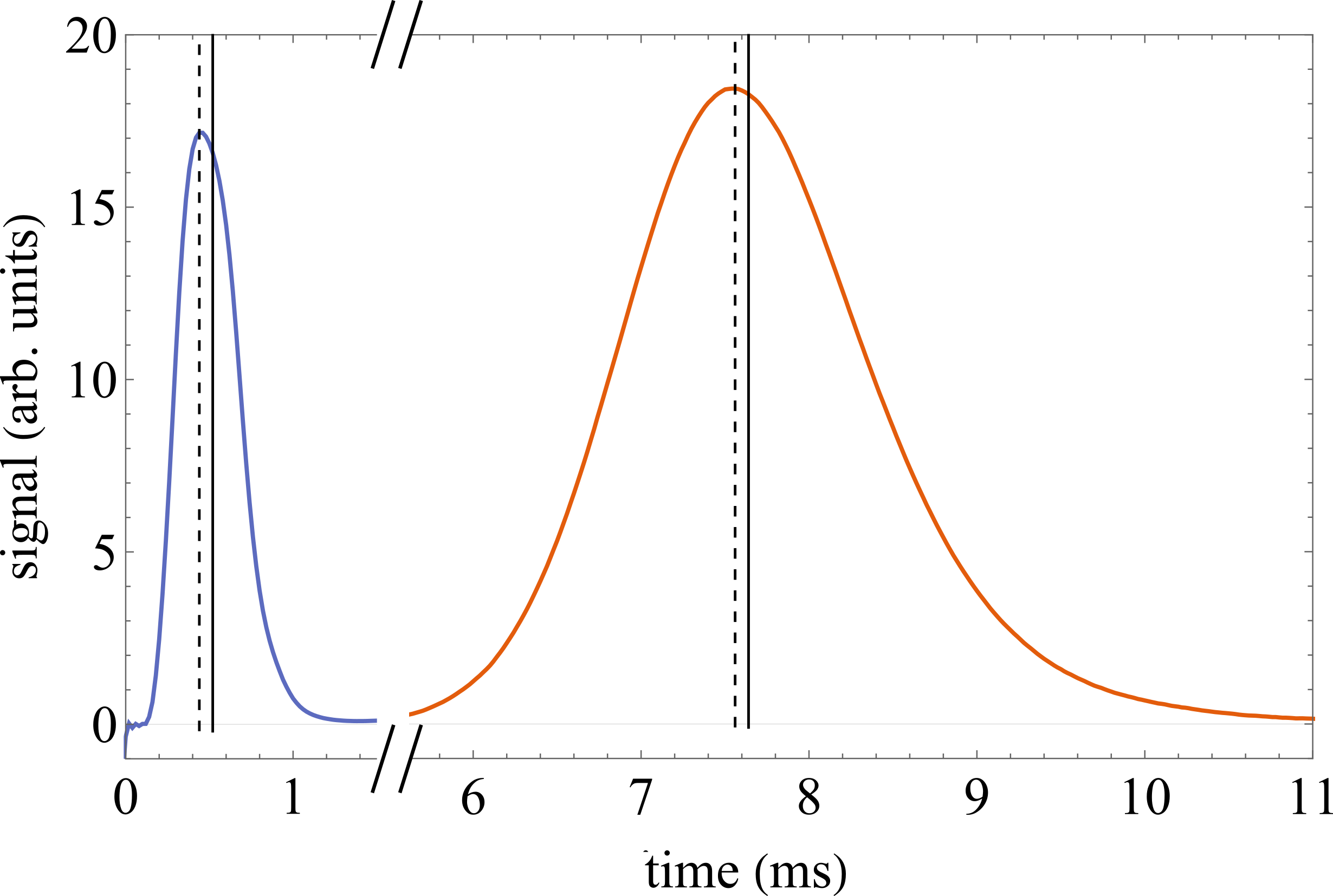}
\caption{Average absorption near the molecule source (blue, left) and fluorescence in the detection region (orange, right) traces show the transit time and the velocity dispersion of the molecular pulses. The peak time (dotted) and average time (solid) differ slightly. The displayed traces are averaged from the entire second data set. The height of the absorption signal has been inverted and re-scaled so it can be compared to the detected fluorescence signal.}
\label{fig:comfl}
\end{center}
\end{figure}

The final source of uncertainty is from the determination of the molecular beam velocity $v$. This is deduced from the time that elapses between when an absorption signal is observed just after the buffer gas cell and the fluorescence signal is observed at a distance $178 \pm 1$ cm away (Fig.~\ref{fig:comfl}). The uncertainty in $\tau_H$ from this uncertainty in the spacing is negligibly small for this large separation. A bigger contribution arises because the molecules do not exit the buffer gas cell at the same time~\cite{hutzlerthesis}\cite{pattersonthesis} and because the velocity dispersion for the molecules changes the shape of the pulse of molecules as they travel this distance. The difference of the average times (solid lines in the figure) for each distribution is used to compute the velocity. The 0.1 ms difference of the peak (dashed lines in Fig.~\ref{fig:comfl}) and average arrival times of each distribution when applied to the dataset results in a 0.05 ms uncertainty in $\tau_H$ from the velocity determination.

The sources of uncertainties in the measurement are summarized in Table~\ref{tab:finalerror}. Assuming these uncertainties are uncorrelated, they together produce the 0.5 ms uncertainty in Eq.~\ref{eq:finresult}. 

\begin{table}[htbp]
\centering
\begin{tabular}{|l|l|}
\hline
\textbf{Source of uncertainty}                  & \textbf{Uncertainty (ms)} \\ \hline
Fitting uncertainty  & 0.02     \\ \hline
Excitation laser saturation  & 0.4                       \\ \hline
Background uncertainty  & 0.2     \\ \hline
Excitation laser position  & 0.04                       \\ \hline
Velocity determination & 0.05 \\ \hline
\textbf{Total uncertainty ($1\sigma$)}          & \textbf{0.5}                       \\ \hline
\end{tabular}
\caption{Table of all systematic and statistical uncertainties of lifetime measurement.}
\label{tab:finalerror}
\end{table}

\section{Results}
\label{sec:results}
 The best fit result for the measured H state lifetime is 
\begin{equation}
    \tau_H = 4.2 \pm 0.5~\mathrm{ms}.
    \label{eq:finresult}
\end{equation}
 The uncertainty stated is a one standard deviation combined statistical and systematic uncertainty. 
As discussed in the last section, the systematic uncertainties dominate.  

We performed three tests to check the robustness of the result and its uncertainty. First, we analyzed the two data sets (orange and blue in Fig.~\ref{fig:twodatafit}) separately, given that they have average molecular velocities that differ by about 20\%, as is clearly evident in Fig. \ref{fig:twodatafit}. Second, we excluded data from one excitation region and refit the remaining data, performing this procedure for each of E2-5. Third, we refit the data while relaxing the requirement that $I_i/I_1 = 1$ at $(L_i-L_1)/v=0$ (see Eq.~\ref{eq:fit}) and converting this expression into a second fit parameter. All of these tests give results for $\tau_H$ that are consistent within the uncertainty.

\section{Implications for a new ACME measurement}
\label{sec:implications}

\begin{figure}[htpb]
\begin{center}
\includegraphics[width=3.25in]{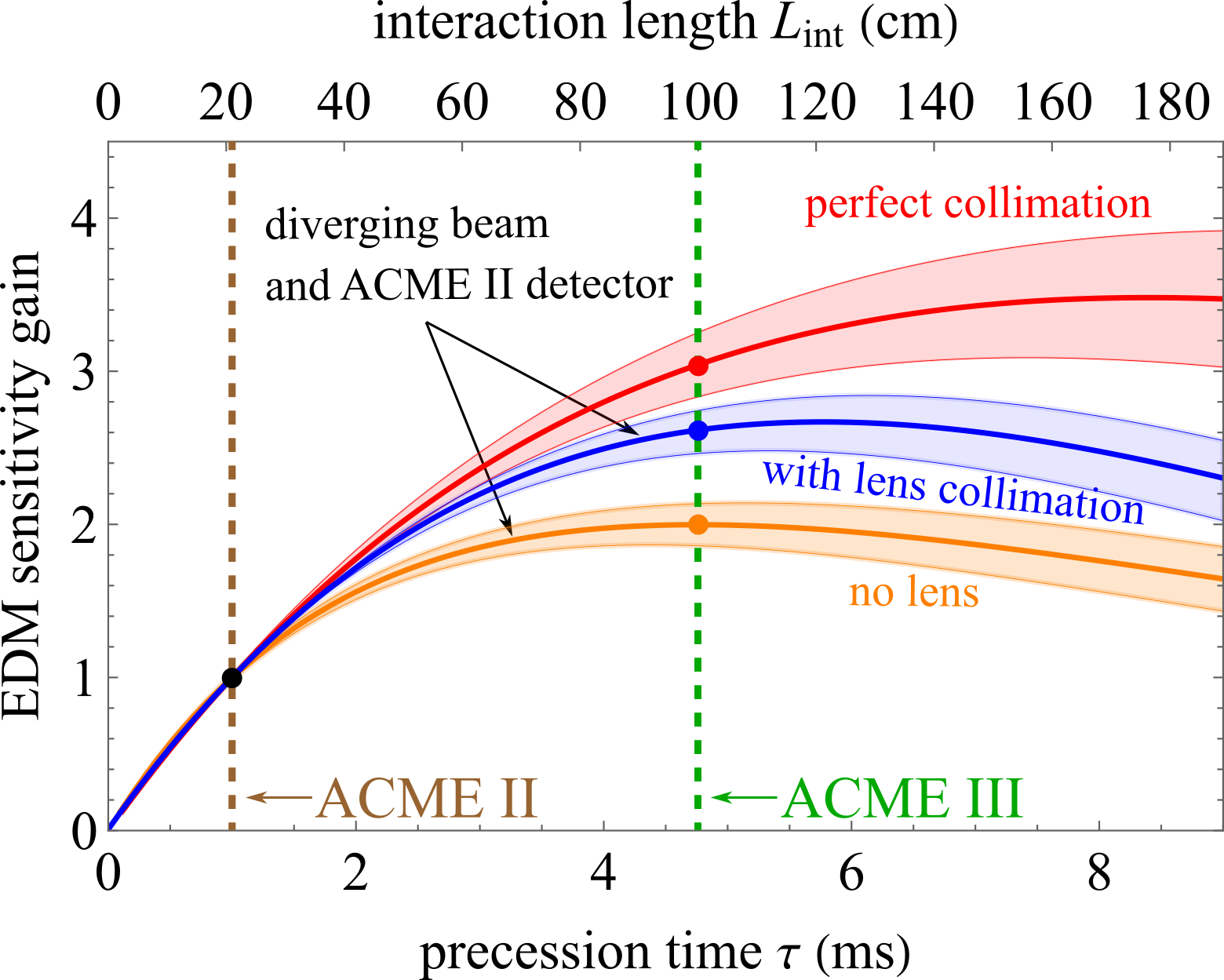}
\caption{Projected EDM sensitivity gains over ACME II given the measured lifetime $\tau_H$ for a perfectly collimated molecular beam (red, upper curve).  The bands represent the effect of the uncertainty in $\tau_H$, and the dashed lines indicate the coherence times for ACME II and the projected ACME III. For a diverging molecular beam in an apparatus that is made longer without increasing its radial dimensions, the longer $\tau_H$ increases the sensitivity by a factor of 2 (orange, lower curve). The sensitivity improves by up to 2.7 due to the effective collimation provided by the addition of an electrostatic lens for the molecules (blue, middle curve).  (The additional sensitivity gain of 3.5 because the lens also captures more molecules is not included.)}
\label{fig:gainedm}
\end{center}
\end{figure}

The new measured value of $\tau_H$ is significantly longer than the coherence time used in the ACME I and II experiments ($\tau\approx1$ ms).  A significant decrease in the uncertainty of the electron EDM $\delta d_e$ (Eq.~\ref{eqn:edmunc}) could thus result from increasing the coherence time $\tau$ in Eq.~\ref{eqn:edmunc} to something close to $\tau_H$. For a limiting case where the molecular beam is perfectly collimated, the optimum $\tau$ for a given $\tau_H$ can be calculated by 
\begin{equation}
    \delta d_e \propto \frac{1}{\tau\sqrt{N}} \propto \frac{1}{\tau \sqrt{\exp{(-\tau/\tau_H)}}}.
    \label{eqn:gainedm}
\end{equation}
The minimum value at $\tau = 2\tau_H$ is $\sim$20\% smaller than when compared to $\tau = \tau_H$. The red, upper curve in Fig.~\ref{fig:gainedm} shows the EDM sensitivity gain relative to ACME II (where $\tau \approx 1$ ms) in this idealized case. The bands in the figure correspond to the uncertainty of this lifetime measurement. A second horizontal axis corresponding to the length of the interaction region $L_{int}$ assuming a beam velocity of $v = 210$ m/s is shown. Extending the coherence time from $1$ ms in ACME II to the planned $\approx5$ ms with a 5 times longer interaction region for ACME III will realize essentially all of this potential gain.

Taking advantage of the longer lifetime for ACME III requires constructing a longer apparatus. Attaining the perfectly collimated limit (red upper curve in Fig.~\ref{fig:gainedm}) with a molecular beam that spreads out, however, would also require a rather impractical scaling up of the radial size of the apparatus.  The size of electric field plates and detection optics would need to increase, as must the power of the lasers needed to saturate the molecules over a larger volume.

It is much more practical to keep the size of the apparatus perpendicular to the molecular beam much the same as for ACME II.  This length increase by itself would then increase the eEDM sensitivity by about a factor of 2 (orange curve in Fig.~\ref{fig:gainedm}).  Simulations suggest that the gain should increase from 2.0 to 2.7 due to the molecule collimation of the electrostatic lens to be used in ACME III (the blue middle curve in Fig.~\ref{fig:gainedm}). The additional sensitivity gain of 3.5 because the lens captures more of the diverging molecules from the ablation source \cite{Wu2020,Wu2022} is not included in the figure.

The longer lifetime and the lens collimation together give a sensitivity gain of 2.7, and the lens capture of more molecules provides an additional factor of 3.5.  SIPM detectors \cite{masudasipm} and upgraded collection optics promise to increase the sensitivity by an additional factor of 2. Reducing excess noise due to a timing imperfection (in ACME II) should increase the statistical sensitivity by a factor of 1.7~\cite{Panda2019}. The projected ACME III sensitivity  over ACME II should be improved by about a factor of 30.   

The ACME I measurement \cite{Baron2014} and ACME II measurement \cite{ACMECollaboration2018} each increased the sensitivity of electron EDM measurements by an order of magnitude.  The significant implication of this lifetime measurement is that a longer ACME III apparatus, with an electrostatic lens and improved detection efficiency, should produce  an additional order of magnitude increase in EDM sensitivity.  

\section*{Acknowledgements}
This work was supported by the National Science Foundation, the Gordon and Betty Moore Foundation, and the Alfred P. Sloan Foundation. D.G.A. was partially supported by the Mustard Seed Foundation and the Amherst
College Kellogg University Fellowship. N. Hutzler, P. Hu, and Z. Han provided useful comments and discussion. This work was performed as part of the ACME collaboration.

\bibliography{ACMEIIbib,ACMEIIIbib}

\begin{thebibliography}{25}%
\makeatletter
\providecommand \@ifxundefined [1]{%
 \@ifx{#1\undefined}
}%
\providecommand \@ifnum [1]{%
 \ifnum #1\expandafter \@firstoftwo
 \else \expandafter \@secondoftwo
 \fi
}%
\providecommand \@ifx [1]{%
 \ifx #1\expandafter \@firstoftwo
 \else \expandafter \@secondoftwo
 \fi
}%
\providecommand \natexlab [1]{#1}%
\providecommand \enquote  [1]{``#1''}%
\providecommand \bibnamefont  [1]{#1}%
\providecommand \bibfnamefont [1]{#1}%
\providecommand \citenamefont [1]{#1}%
\providecommand \href@noop [0]{\@secondoftwo}%
\providecommand \href [0]{\begingroup \@sanitize@url \@href}%
\providecommand \@href[1]{\@@startlink{#1}\@@href}%
\providecommand \@@href[1]{\endgroup#1\@@endlink}%
\providecommand \@sanitize@url [0]{\catcode `\\12\catcode `\$12\catcode
  `\&12\catcode `\#12\catcode `\^12\catcode `\_12\catcode `\%12\relax}%
\providecommand \@@startlink[1]{}%
\providecommand \@@endlink[0]{}%
\providecommand \url  [0]{\begingroup\@sanitize@url \@url }%
\providecommand \@url [1]{\endgroup\@href {#1}{\urlprefix }}%
\providecommand \urlprefix  [0]{URL }%
\providecommand \Eprint [0]{\href }%
\providecommand \doibase [0]{https://doi.org/}%
\providecommand \selectlanguage [0]{\@gobble}%
\providecommand \bibinfo  [0]{\@secondoftwo}%
\providecommand \bibfield  [0]{\@secondoftwo}%
\providecommand \translation [1]{[#1]}%
\providecommand \BibitemOpen [0]{}%
\providecommand \bibitemStop [0]{}%
\providecommand \bibitemNoStop [0]{.\EOS\space}%
\providecommand \EOS [0]{\spacefactor3000\relax}%
\providecommand \BibitemShut  [1]{\csname bibitem#1\endcsname}%
\let\auto@bib@innerbib\@empty
\bibitem [{\citenamefont {{ACME Collaboration}}\ \emph
  {et~al.}(2018)\citenamefont {{ACME Collaboration}}, \citenamefont {Andreev},
  \citenamefont {Ang}, \citenamefont {DeMille}, \citenamefont {Doyle},
  \citenamefont {Gabrielse}, \citenamefont {Haefner}, \citenamefont {Hutzler},
  \citenamefont {Lasner}, \citenamefont {Meisenhelder}, \citenamefont
  {O'Leary}, \citenamefont {Panda}, \citenamefont {West}, \citenamefont
  {West},\ and\ \citenamefont {Wu}}]{ACMECollaboration2018}%
  \BibitemOpen
  \bibfield  {author} {\bibinfo {author} {\bibnamefont {{ACME Collaboration}}},
  \bibinfo {author} {\bibfnamefont {V.}~\bibnamefont {Andreev}}, \bibinfo
  {author} {\bibfnamefont {D.~G.}\ \bibnamefont {Ang}}, \bibinfo {author}
  {\bibfnamefont {D.}~\bibnamefont {DeMille}}, \bibinfo {author} {\bibfnamefont
  {J.~M.}\ \bibnamefont {Doyle}}, \bibinfo {author} {\bibfnamefont
  {G.}~\bibnamefont {Gabrielse}}, \bibinfo {author} {\bibfnamefont
  {J.}~\bibnamefont {Haefner}}, \bibinfo {author} {\bibfnamefont {N.~R.}\
  \bibnamefont {Hutzler}}, \bibinfo {author} {\bibfnamefont {Z.}~\bibnamefont
  {Lasner}}, \bibinfo {author} {\bibfnamefont {C.}~\bibnamefont
  {Meisenhelder}}, \bibinfo {author} {\bibfnamefont {B.~R.}\ \bibnamefont
  {O'Leary}}, \bibinfo {author} {\bibfnamefont {C.~D.}\ \bibnamefont {Panda}},
  \bibinfo {author} {\bibfnamefont {A.~D.}\ \bibnamefont {West}}, \bibinfo
  {author} {\bibfnamefont {E.~P.}\ \bibnamefont {West}},\ and\ \bibinfo
  {author} {\bibfnamefont {X.}~\bibnamefont {Wu}},\ }\bibfield  {title}
  {\bibinfo {title} {{Improved limit on the electric dipole moment of the
  electron}},\ }\href {https://doi.org/10.1038/s41586-018-0599-8} {\bibfield
  {journal} {\bibinfo  {journal} {Nature}\ }\textbf {\bibinfo {volume} {562}},\
  \bibinfo {pages} {355} (\bibinfo {year} {2018})}\BibitemShut {NoStop}%
\bibitem [{\citenamefont {Baron}\ \emph {et~al.}(2014)\citenamefont {Baron},
  \citenamefont {Campbell}, \citenamefont {DeMille}, \citenamefont {Doyle},
  \citenamefont {Gabrielse}, \citenamefont {Gurevich}, \citenamefont {Hess},
  \citenamefont {Hutzler}, \citenamefont {Kirilov}, \citenamefont {Kozyryev},
  \citenamefont {O'Leary}, \citenamefont {Panda}, \citenamefont {Parsons},
  \citenamefont {Petrik}, \citenamefont {Spaun}, \citenamefont {Vutha},\ and\
  \citenamefont {West}}]{Baron2014}%
  \BibitemOpen
  \bibfield  {author} {\bibinfo {author} {\bibfnamefont {J.}~\bibnamefont
  {Baron}}, \bibinfo {author} {\bibfnamefont {W.~C.}\ \bibnamefont {Campbell}},
  \bibinfo {author} {\bibfnamefont {D.}~\bibnamefont {DeMille}}, \bibinfo
  {author} {\bibfnamefont {J.~M.}\ \bibnamefont {Doyle}}, \bibinfo {author}
  {\bibfnamefont {G.}~\bibnamefont {Gabrielse}}, \bibinfo {author}
  {\bibfnamefont {Y.~V.}\ \bibnamefont {Gurevich}}, \bibinfo {author}
  {\bibfnamefont {P.~W.}\ \bibnamefont {Hess}}, \bibinfo {author}
  {\bibfnamefont {N.~R.}\ \bibnamefont {Hutzler}}, \bibinfo {author}
  {\bibfnamefont {E.}~\bibnamefont {Kirilov}}, \bibinfo {author} {\bibfnamefont
  {I.}~\bibnamefont {Kozyryev}}, \bibinfo {author} {\bibfnamefont {B.~R.}\
  \bibnamefont {O'Leary}}, \bibinfo {author} {\bibfnamefont {C.~D.}\
  \bibnamefont {Panda}}, \bibinfo {author} {\bibfnamefont {M.~F.}\ \bibnamefont
  {Parsons}}, \bibinfo {author} {\bibfnamefont {E.~S.}\ \bibnamefont {Petrik}},
  \bibinfo {author} {\bibfnamefont {B.}~\bibnamefont {Spaun}}, \bibinfo
  {author} {\bibfnamefont {A.~C.}\ \bibnamefont {Vutha}},\ and\ \bibinfo
  {author} {\bibfnamefont {A.~D.}\ \bibnamefont {West}},\ }\bibfield  {title}
  {\bibinfo {title} {{Order of magnitude smaller limit on the electric dipole
  moment of the electron.}},\ }\href {https://doi.org/10.1126/science.1248213}
  {\bibfield  {journal} {\bibinfo  {journal} {Science}\ }\textbf {\bibinfo
  {volume} {343}},\ \bibinfo {pages} {269} (\bibinfo {year}
  {2014})}\BibitemShut {NoStop}%
\bibitem [{\citenamefont {Cairncross}\ \emph {et~al.}(2017)\citenamefont
  {Cairncross}, \citenamefont {Gresh}, \citenamefont {Grau}, \citenamefont
  {Cossel}, \citenamefont {Roussy}, \citenamefont {Ni}, \citenamefont {Zhou},
  \citenamefont {Ye},\ and\ \citenamefont {Cornell}}]{Cairncross2017}%
  \BibitemOpen
  \bibfield  {author} {\bibinfo {author} {\bibfnamefont {W.~B.}\ \bibnamefont
  {Cairncross}}, \bibinfo {author} {\bibfnamefont {D.~N.}\ \bibnamefont
  {Gresh}}, \bibinfo {author} {\bibfnamefont {M.}~\bibnamefont {Grau}},
  \bibinfo {author} {\bibfnamefont {K.~C.}\ \bibnamefont {Cossel}}, \bibinfo
  {author} {\bibfnamefont {T.~S.}\ \bibnamefont {Roussy}}, \bibinfo {author}
  {\bibfnamefont {Y.}~\bibnamefont {Ni}}, \bibinfo {author} {\bibfnamefont
  {Y.}~\bibnamefont {Zhou}}, \bibinfo {author} {\bibfnamefont {J.}~\bibnamefont
  {Ye}},\ and\ \bibinfo {author} {\bibfnamefont {E.~A.}\ \bibnamefont
  {Cornell}},\ }\bibfield  {title} {\bibinfo {title} {Precision measurement of
  the electron's electric dipole moment using trapped molecular ions},\ }\href
  {https://doi.org/10.1103/PhysRevLett.119.153001} {\bibfield  {journal}
  {\bibinfo  {journal} {Phys. Rev. Lett.}\ }\textbf {\bibinfo {volume} {119}},\
  \bibinfo {pages} {153001} (\bibinfo {year} {2017})}\BibitemShut {NoStop}%
\bibitem [{\citenamefont {Cesarotti}\ \emph {et~al.}(2019)\citenamefont
  {Cesarotti}, \citenamefont {Lu}, \citenamefont {Nakai}, \citenamefont
  {Parikh},\ and\ \citenamefont {Reece}}]{Cesarotti2019}%
  \BibitemOpen
  \bibfield  {author} {\bibinfo {author} {\bibfnamefont {C.}~\bibnamefont
  {Cesarotti}}, \bibinfo {author} {\bibfnamefont {Q.}~\bibnamefont {Lu}},
  \bibinfo {author} {\bibfnamefont {Y.}~\bibnamefont {Nakai}}, \bibinfo
  {author} {\bibfnamefont {A.}~\bibnamefont {Parikh}},\ and\ \bibinfo {author}
  {\bibfnamefont {M.}~\bibnamefont {Reece}},\ }\bibfield  {title} {\bibinfo
  {title} {Interpreting the electron edm constraint},\ }\href
  {https://doi.org/10.1007/JHEP05(2019)059} {\bibfield  {journal} {\bibinfo
  {journal} {Journal of High Energy Physics}\ }\textbf {\bibinfo {volume}
  {2019}},\ \bibinfo {pages} {59} (\bibinfo {year} {2019})}\BibitemShut
  {NoStop}%
\bibitem [{\citenamefont {Panico}\ \emph {et~al.}(2019)\citenamefont {Panico},
  \citenamefont {Pomarol},\ and\ \citenamefont {Riembau}}]{Panico2019}%
  \BibitemOpen
  \bibfield  {author} {\bibinfo {author} {\bibfnamefont {G.}~\bibnamefont
  {Panico}}, \bibinfo {author} {\bibfnamefont {A.}~\bibnamefont {Pomarol}},\
  and\ \bibinfo {author} {\bibfnamefont {M.}~\bibnamefont {Riembau}},\
  }\bibfield  {title} {\bibinfo {title} {Eft approach to the electron electric
  dipole moment at the two-loop level},\ }\href
  {https://doi.org/10.1007/JHEP04(2019)090} {\bibfield  {journal} {\bibinfo
  {journal} {Journal of High Energy Physics}\ }\textbf {\bibinfo {volume}
  {2019}},\ \bibinfo {pages} {90} (\bibinfo {year} {2019})}\BibitemShut
  {NoStop}%
\bibitem [{\citenamefont {Denis}\ and\ \citenamefont
  {Fleig}(2016)}]{Denis2016}%
  \BibitemOpen
  \bibfield  {author} {\bibinfo {author} {\bibfnamefont {M.}~\bibnamefont
  {Denis}}\ and\ \bibinfo {author} {\bibfnamefont {T.}~\bibnamefont {Fleig}},\
  }\bibfield  {title} {\bibinfo {title} {{In search of discrete symmetry
  violations beyond the standard model: Thorium monoxide reloaded}},\
  }\bibfield  {journal} {\bibinfo  {journal} {J. Chem. Phys.}\ }\textbf
  {\bibinfo {volume} {145}},\ \href {https://doi.org/10.1063/1.4968597}
  {10.1063/1.4968597} (\bibinfo {year} {2016})\BibitemShut {NoStop}%
\bibitem [{\citenamefont {Skripnikov}(2016)}]{Skripnikov2016}%
  \BibitemOpen
  \bibfield  {author} {\bibinfo {author} {\bibfnamefont {L.~V.}\ \bibnamefont
  {Skripnikov}},\ }\bibfield  {title} {\bibinfo {title} {{Combined 4-component
  and relativistic pseudopotential study of ThO for the electron electric
  dipole moment search}},\ }\bibfield  {journal} {\bibinfo  {journal} {J .Chem.
  Phys.}\ }\textbf {\bibinfo {volume} {145}},\ \href
  {https://doi.org/10.1063/1.4968229} {10.1063/1.4968229} (\bibinfo {year}
  {2016})\BibitemShut {NoStop}%
\bibitem [{\citenamefont {Kirilov}\ \emph {et~al.}(2013)\citenamefont
  {Kirilov}, \citenamefont {Campbell}, \citenamefont {Doyle}, \citenamefont
  {Gabrielse}, \citenamefont {Gurevich}, \citenamefont {Hess}, \citenamefont
  {Hutzler}, \citenamefont {O'Leary}, \citenamefont {Petrik}, \citenamefont
  {Spaun}, \citenamefont {Vutha},\ and\ \citenamefont {DeMille}}]{Kirilov2013}%
  \BibitemOpen
  \bibfield  {author} {\bibinfo {author} {\bibfnamefont {E.}~\bibnamefont
  {Kirilov}}, \bibinfo {author} {\bibfnamefont {W.~C.}\ \bibnamefont
  {Campbell}}, \bibinfo {author} {\bibfnamefont {J.~M.}\ \bibnamefont {Doyle}},
  \bibinfo {author} {\bibfnamefont {G.}~\bibnamefont {Gabrielse}}, \bibinfo
  {author} {\bibfnamefont {Y.~V.}\ \bibnamefont {Gurevich}}, \bibinfo {author}
  {\bibfnamefont {P.~W.}\ \bibnamefont {Hess}}, \bibinfo {author}
  {\bibfnamefont {N.~R.}\ \bibnamefont {Hutzler}}, \bibinfo {author}
  {\bibfnamefont {B.~R.}\ \bibnamefont {O'Leary}}, \bibinfo {author}
  {\bibfnamefont {E.}~\bibnamefont {Petrik}}, \bibinfo {author} {\bibfnamefont
  {B.}~\bibnamefont {Spaun}}, \bibinfo {author} {\bibfnamefont {A.~C.}\
  \bibnamefont {Vutha}},\ and\ \bibinfo {author} {\bibfnamefont
  {D.}~\bibnamefont {DeMille}},\ }\bibfield  {title} {\bibinfo {title}
  {{Shot-noise-limited spin measurements in a pulsed molecular beam}},\ }\href
  {https://doi.org/10.1103/PhysRevA.88.013844} {\bibfield  {journal} {\bibinfo
  {journal} {Phys. Rev. A}\ }\textbf {\bibinfo {volume} {88}},\ \bibinfo
  {pages} {013844} (\bibinfo {year} {2013})}\BibitemShut {NoStop}%
\bibitem [{\citenamefont {Panda}\ \emph {et~al.}(2019)\citenamefont {Panda},
  \citenamefont {Meisenhelder}, \citenamefont {Verma}, \citenamefont {Ang},
  \citenamefont {Chow}, \citenamefont {Lasner}, \citenamefont {Wu},
  \citenamefont {DeMille}, \citenamefont {Doyle},\ and\ \citenamefont
  {Gabrielse}}]{Panda2019}%
  \BibitemOpen
  \bibfield  {author} {\bibinfo {author} {\bibfnamefont {C.~D.}\ \bibnamefont
  {Panda}}, \bibinfo {author} {\bibfnamefont {C.}~\bibnamefont {Meisenhelder}},
  \bibinfo {author} {\bibfnamefont {M.}~\bibnamefont {Verma}}, \bibinfo
  {author} {\bibfnamefont {D.~G.}\ \bibnamefont {Ang}}, \bibinfo {author}
  {\bibfnamefont {J.}~\bibnamefont {Chow}}, \bibinfo {author} {\bibfnamefont
  {Z.}~\bibnamefont {Lasner}}, \bibinfo {author} {\bibfnamefont
  {X.}~\bibnamefont {Wu}}, \bibinfo {author} {\bibfnamefont {D.}~\bibnamefont
  {DeMille}}, \bibinfo {author} {\bibfnamefont {J.~M.}\ \bibnamefont {Doyle}},\
  and\ \bibinfo {author} {\bibfnamefont {G.}~\bibnamefont {Gabrielse}},\
  }\bibfield  {title} {\bibinfo {title} {Attaining the shot-noise-limit in the
  {ACME} measurement of the electron electric dipole moment},\ }\href
  {https://doi.org/10.1088/1361-6455/ab4a61} {\bibfield  {journal} {\bibinfo
  {journal} {Journal of Physics B: Atomic, Molecular and Optical Physics}\
  }\textbf {\bibinfo {volume} {52}},\ \bibinfo {pages} {235003} (\bibinfo
  {year} {2019})}\BibitemShut {NoStop}%
\bibitem [{\citenamefont {Vutha}\ \emph {et~al.}(2010)\citenamefont {Vutha},
  \citenamefont {Campbell}, \citenamefont {Gurevich}, \citenamefont {Hutzler},
  \citenamefont {Parsons}, \citenamefont {Patterson}, \citenamefont {Petrik},
  \citenamefont {Spaun}, \citenamefont {Doyle}, \citenamefont {Gabrielse},\
  and\ \citenamefont {DeMille}}]{Vutha2010}%
  \BibitemOpen
  \bibfield  {author} {\bibinfo {author} {\bibfnamefont {A.~C.}\ \bibnamefont
  {Vutha}}, \bibinfo {author} {\bibfnamefont {W.~C.}\ \bibnamefont {Campbell}},
  \bibinfo {author} {\bibfnamefont {Y.~V.}\ \bibnamefont {Gurevich}}, \bibinfo
  {author} {\bibfnamefont {N.~R.}\ \bibnamefont {Hutzler}}, \bibinfo {author}
  {\bibfnamefont {M.}~\bibnamefont {Parsons}}, \bibinfo {author} {\bibfnamefont
  {D.}~\bibnamefont {Patterson}}, \bibinfo {author} {\bibfnamefont
  {E.}~\bibnamefont {Petrik}}, \bibinfo {author} {\bibfnamefont
  {B.}~\bibnamefont {Spaun}}, \bibinfo {author} {\bibfnamefont {J.~M.}\
  \bibnamefont {Doyle}}, \bibinfo {author} {\bibfnamefont {G.}~\bibnamefont
  {Gabrielse}},\ and\ \bibinfo {author} {\bibfnamefont {D.}~\bibnamefont
  {DeMille}},\ }\bibfield  {title} {\bibinfo {title} {{Search for the electric
  dipole moment of the electron with thorium monoxide}},\ }\bibfield  {journal}
  {\bibinfo  {journal} {J. Phys. B}\ }\textbf {\bibinfo {volume} {43}},\ \href
  {https://doi.org/10.1088/0953-4075/44/7/079803}
  {10.1088/0953-4075/44/7/079803} (\bibinfo {year} {2010})\BibitemShut
  {NoStop}%
\bibitem [{\citenamefont {Vutha}(2011)}]{vuthathesis}%
  \BibitemOpen
  \bibfield  {author} {\bibinfo {author} {\bibfnamefont {A.~C.}\ \bibnamefont
  {Vutha}},\ }\emph {\bibinfo {title} {A search for the electric dipole moment
  of the electron using thorium monoxide}},\ \href@noop {} {Ph.D. thesis},\
  \bibinfo  {school} {Yale University} (\bibinfo {year} {2011})\BibitemShut
  {NoStop}%
\bibitem [{\citenamefont {Au}(2014)}]{yatthesis}%
  \BibitemOpen
  \bibfield  {author} {\bibinfo {author} {\bibfnamefont {Y.~S.}\ \bibnamefont
  {Au}},\ }\emph {\bibinfo {title} {Inelastic collisions of atomic thorium and
  molecular thorium monoxide with cold helium-3}},\ \href@noop {} {Ph.D.
  thesis},\ \bibinfo  {school} {Harvard University} (\bibinfo {year}
  {2014})\BibitemShut {NoStop}%
\bibitem [{\citenamefont {Hutzler}\ \emph {et~al.}(2011)\citenamefont
  {Hutzler}, \citenamefont {Parsons}, \citenamefont {Gurevich}, \citenamefont
  {Hess}, \citenamefont {Petrik}, \citenamefont {Spaun}, \citenamefont {Vutha},
  \citenamefont {DeMille}, \citenamefont {Gabrielse},\ and\ \citenamefont
  {Doyle}}]{Hutzler2011a}%
  \BibitemOpen
  \bibfield  {author} {\bibinfo {author} {\bibfnamefont {N.~R.}\ \bibnamefont
  {Hutzler}}, \bibinfo {author} {\bibfnamefont {M.~F.}\ \bibnamefont
  {Parsons}}, \bibinfo {author} {\bibfnamefont {Y.~V.}\ \bibnamefont
  {Gurevich}}, \bibinfo {author} {\bibfnamefont {P.~W.}\ \bibnamefont {Hess}},
  \bibinfo {author} {\bibfnamefont {E.}~\bibnamefont {Petrik}}, \bibinfo
  {author} {\bibfnamefont {B.}~\bibnamefont {Spaun}}, \bibinfo {author}
  {\bibfnamefont {A.~C.}\ \bibnamefont {Vutha}}, \bibinfo {author}
  {\bibfnamefont {D.}~\bibnamefont {DeMille}}, \bibinfo {author} {\bibfnamefont
  {G.}~\bibnamefont {Gabrielse}},\ and\ \bibinfo {author} {\bibfnamefont
  {J.~M.}\ \bibnamefont {Doyle}},\ }\bibfield  {title} {\bibinfo {title} {{A
  cryogenic beam of refractory, chemically reactive molecules with expansion
  cooling}},\ }\href {https://doi.org/10.1039/c1cp20901a} {\bibfield  {journal}
  {\bibinfo  {journal} {Phys. Chem. Chem. Phys.}\ }\textbf {\bibinfo {volume}
  {13}},\ \bibinfo {pages} {18976} (\bibinfo {year} {2011})}\BibitemShut
  {NoStop}%
\bibitem [{\citenamefont {Spaun}(2014)}]{spaun}%
  \BibitemOpen
  \bibfield  {author} {\bibinfo {author} {\bibfnamefont {B.~N.}\ \bibnamefont
  {Spaun}},\ }\emph {\bibinfo {title} {A Ten-Fold Improvement to the Limit of
  the Electron Electric Dipole Moment}},\ \href@noop {} {Ph.D. thesis},\
  \bibinfo  {school} {Harvard University} (\bibinfo {year} {2014})\BibitemShut
  {NoStop}%
\bibitem [{\citenamefont {Sch{\"u}nemann}\ \emph {et~al.}(1999)\citenamefont
  {Sch{\"u}nemann}, \citenamefont {Engler}, \citenamefont {Grimm},
  \citenamefont {Weidem{\"a}ller},\ and\ \citenamefont
  {Zielonkowski}}]{delaylinelock}%
  \BibitemOpen
  \bibfield  {author} {\bibinfo {author} {\bibfnamefont {U.}~\bibnamefont
  {Sch{\"u}nemann}}, \bibinfo {author} {\bibfnamefont {H.}~\bibnamefont
  {Engler}}, \bibinfo {author} {\bibfnamefont {R.}~\bibnamefont {Grimm}},
  \bibinfo {author} {\bibfnamefont {M.}~\bibnamefont {Weidem{\"a}ller}},\ and\
  \bibinfo {author} {\bibfnamefont {M.}~\bibnamefont {Zielonkowski}},\
  }\bibfield  {title} {\bibinfo {title} {Simple scheme for tunable frequency
  offset locking of two lasers},\ }\href {https://doi.org/10.1063/1.1149573}
  {\bibfield  {journal} {\bibinfo  {journal} {Review of Scientific
  Instruments}\ }\textbf {\bibinfo {volume} {70}},\ \bibinfo {pages} {242}
  (\bibinfo {year} {1999})},\ \Eprint
  {https://arxiv.org/abs/https://doi.org/10.1063/1.1149573}
  {https://doi.org/10.1063/1.1149573} \BibitemShut {NoStop}%
\bibitem [{\citenamefont {Panda}(2019)}]{pandathesis}%
  \BibitemOpen
  \bibfield  {author} {\bibinfo {author} {\bibfnamefont {C.~D.}\ \bibnamefont
  {Panda}},\ }\emph {\bibinfo {title} {Order of Magnitude Improved Limit on the
  Electric Dipole Moment of the Electron}},\ \href@noop {} {Ph.D. thesis},\
  \bibinfo  {school} {Harvard University} (\bibinfo {year} {2019})\BibitemShut
  {NoStop}%
\bibitem [{\citenamefont {Drever}\ \emph {et~al.}(1983)\citenamefont {Drever},
  \citenamefont {Hall}, \citenamefont {Kowalski}, \citenamefont {Hough},
  \citenamefont {Ford}, \citenamefont {Munley},\ and\ \citenamefont
  {Ward}}]{Drever}%
  \BibitemOpen
  \bibfield  {author} {\bibinfo {author} {\bibfnamefont {R.~W.~P.}\
  \bibnamefont {Drever}}, \bibinfo {author} {\bibfnamefont {J.~L.}\
  \bibnamefont {Hall}}, \bibinfo {author} {\bibfnamefont {F.~V.}\ \bibnamefont
  {Kowalski}}, \bibinfo {author} {\bibfnamefont {J.}~\bibnamefont {Hough}},
  \bibinfo {author} {\bibfnamefont {G.~M.}\ \bibnamefont {Ford}}, \bibinfo
  {author} {\bibfnamefont {A.~J.}\ \bibnamefont {Munley}},\ and\ \bibinfo
  {author} {\bibfnamefont {H.}~\bibnamefont {Ward}},\ }\bibfield  {title}
  {\bibinfo {title} {Laser phase and frequency stabilization using an optical
  resonator},\ }\href {https://doi.org/10.1007/BF00702605} {\bibfield
  {journal} {\bibinfo  {journal} {Applied Physics B}\ }\textbf {\bibinfo
  {volume} {31}},\ \bibinfo {pages} {97} (\bibinfo {year} {1983})}\BibitemShut
  {NoStop}%
\bibitem [{\citenamefont {{Baron}}\ \emph {et~al.}(2017)\citenamefont
  {{Baron}}, \citenamefont {{Campbell}}, \citenamefont {{DeMille}},
  \citenamefont {{Doyle}}, \citenamefont {{Gabrielse}}, \citenamefont
  {{Gurevich}}, \citenamefont {{Hess}}, \citenamefont {{Hutzler}},
  \citenamefont {{Kirilov}}, \citenamefont {{Kozyryev}}, \citenamefont
  {{O'Leary}}, \citenamefont {{Panda}}, \citenamefont {{Parsons}},
  \citenamefont {{Spaun}}, \citenamefont {{Vutha}}, \citenamefont {{West}},
  \citenamefont {{West}},\ and\ \citenamefont {{ACME
  Collaboration}}}]{Baron2017}%
  \BibitemOpen
  \bibfield  {author} {\bibinfo {author} {\bibfnamefont {J.}~\bibnamefont
  {{Baron}}}, \bibinfo {author} {\bibfnamefont {W.~C.}\ \bibnamefont
  {{Campbell}}}, \bibinfo {author} {\bibfnamefont {D.}~\bibnamefont
  {{DeMille}}}, \bibinfo {author} {\bibfnamefont {J.~M.}\ \bibnamefont
  {{Doyle}}}, \bibinfo {author} {\bibfnamefont {G.}~\bibnamefont
  {{Gabrielse}}}, \bibinfo {author} {\bibfnamefont {Y.~V.}\ \bibnamefont
  {{Gurevich}}}, \bibinfo {author} {\bibfnamefont {P.~W.}\ \bibnamefont
  {{Hess}}}, \bibinfo {author} {\bibfnamefont {N.~R.}\ \bibnamefont
  {{Hutzler}}}, \bibinfo {author} {\bibfnamefont {E.}~\bibnamefont
  {{Kirilov}}}, \bibinfo {author} {\bibfnamefont {I.}~\bibnamefont
  {{Kozyryev}}}, \bibinfo {author} {\bibfnamefont {B.~R.}\ \bibnamefont
  {{O'Leary}}}, \bibinfo {author} {\bibfnamefont {C.~D.}\ \bibnamefont
  {{Panda}}}, \bibinfo {author} {\bibfnamefont {M.~F.}\ \bibnamefont
  {{Parsons}}}, \bibinfo {author} {\bibfnamefont {B.}~\bibnamefont {{Spaun}}},
  \bibinfo {author} {\bibfnamefont {A.~C.}\ \bibnamefont {{Vutha}}}, \bibinfo
  {author} {\bibfnamefont {A.~D.}\ \bibnamefont {{West}}}, \bibinfo {author}
  {\bibfnamefont {E.~P.}\ \bibnamefont {{West}}},\ and\ \bibinfo {author}
  {\bibnamefont {{ACME Collaboration}}},\ }\bibfield  {title} {\bibinfo {title}
  {{Methods, analysis, and the treatment of systematic errors for the electron
  electric dipole moment search in thorium monoxide}},\ }\href
  {https://doi.org/10.1088/1367-2630/aa708e} {\bibfield  {journal} {\bibinfo
  {journal} {New Journal of Physics}\ }\textbf {\bibinfo {volume} {19}},\
  \bibinfo {eid} {073029} (\bibinfo {year} {2017})},\ \Eprint
  {https://arxiv.org/abs/1612.09318} {arXiv:1612.09318 [physics.atom-ph]}
  \BibitemShut {NoStop}%
\bibitem [{\citenamefont {Lyons}(1989)}]{lyons1989statistics}%
  \BibitemOpen
  \bibfield  {author} {\bibinfo {author} {\bibfnamefont {L.}~\bibnamefont
  {Lyons}},\ }\href {https://books.google.com/books?id=C1Vsbk5UkBAC} {\emph
  {\bibinfo {title} {Statistics for Nuclear and Particle Physicists}}}\
  (\bibinfo  {publisher} {Cambridge University Press},\ \bibinfo {year}
  {1989})\BibitemShut {NoStop}%
\bibitem [{\citenamefont {Efron}\ and\ \citenamefont
  {Tibshirani}(1986)}]{Efron1986}%
  \BibitemOpen
  \bibfield  {author} {\bibinfo {author} {\bibfnamefont {B.}~\bibnamefont
  {Efron}}\ and\ \bibinfo {author} {\bibfnamefont {R.}~\bibnamefont
  {Tibshirani}},\ }\bibfield  {title} {\bibinfo {title} {{Bootstrap Methods for
  Standard Errors, Confidence Intervals, and Other Measures of Statistical
  Accuracy}},\ }\href {https://doi.org/10.1214/ss/1177013815} {\bibfield
  {journal} {\bibinfo  {journal} {Stat. Sci.}\ }\textbf {\bibinfo {volume}
  {1}},\ \bibinfo {pages} {54} (\bibinfo {year} {1986})}\BibitemShut {NoStop}%
\bibitem [{\citenamefont {Hutzler}(2014)}]{hutzlerthesis}%
  \BibitemOpen
  \bibfield  {author} {\bibinfo {author} {\bibfnamefont {N.~R.}\ \bibnamefont
  {Hutzler}},\ }\emph {\bibinfo {title} {A New Limit on the Electron Electric
  Dipole Moment: Beam Production, Data Interpretation, and Systematics}},\
  \href@noop {} {Ph.D. thesis},\ \bibinfo  {school} {Harvard University}
  (\bibinfo {year} {2014})\BibitemShut {NoStop}%
\bibitem [{\citenamefont {Patterson}(2010)}]{pattersonthesis}%
  \BibitemOpen
  \bibfield  {author} {\bibinfo {author} {\bibfnamefont {D.}~\bibnamefont
  {Patterson}},\ }\emph {\bibinfo {title} {Buffer Gas Cooled Beams and Cold
  Molecular Collisions}},\ \href@noop {} {Ph.D. thesis},\ \bibinfo  {school}
  {Harvard University} (\bibinfo {year} {2010})\BibitemShut {NoStop}%
\bibitem [{\citenamefont {Wu}\ \emph {et~al.}(2020)\citenamefont {Wu},
  \citenamefont {Han}, \citenamefont {Chow}, \citenamefont {Ang}, \citenamefont
  {Meisenhelder}, \citenamefont {Panda}, \citenamefont {West}, \citenamefont
  {Gabrielse}, \citenamefont {Doyle},\ and\ \citenamefont {DeMille}}]{Wu2020}%
  \BibitemOpen
  \bibfield  {author} {\bibinfo {author} {\bibfnamefont {X.}~\bibnamefont
  {Wu}}, \bibinfo {author} {\bibfnamefont {Z.}~\bibnamefont {Han}}, \bibinfo
  {author} {\bibfnamefont {J.}~\bibnamefont {Chow}}, \bibinfo {author}
  {\bibfnamefont {D.~G.}\ \bibnamefont {Ang}}, \bibinfo {author} {\bibfnamefont
  {C.}~\bibnamefont {Meisenhelder}}, \bibinfo {author} {\bibfnamefont {C.~D.}\
  \bibnamefont {Panda}}, \bibinfo {author} {\bibfnamefont {E.~P.}\ \bibnamefont
  {West}}, \bibinfo {author} {\bibfnamefont {G.}~\bibnamefont {Gabrielse}},
  \bibinfo {author} {\bibfnamefont {J.~M.}\ \bibnamefont {Doyle}},\ and\
  \bibinfo {author} {\bibfnamefont {D.}~\bibnamefont {DeMille}},\ }\bibfield
  {title} {\bibinfo {title} {The metastable {Q $^3\Delta_2$} state of {ThO}: a
  new resource for the {ACME} electron {EDM} search},\ }\href
  {https://doi.org/10.1088/1367-2630/ab6a3a} {\bibfield  {journal} {\bibinfo
  {journal} {New Journal of Physics}\ }\textbf {\bibinfo {volume} {22}},\
  \bibinfo {pages} {023013} (\bibinfo {year} {2020})}\BibitemShut {NoStop}%
\bibitem [{\citenamefont {{Wu}}\ \emph {et~al.}()\citenamefont {{Wu}},
  \citenamefont {{Hu}}, \citenamefont {{Han}}, \citenamefont {{Ang}},
  \citenamefont {{Meisenhelder}}, \citenamefont {{Gabrielse}}, \citenamefont
  {{Doyle}},\ and\ \citenamefont {{DeMille}}}]{Wu2022}%
  \BibitemOpen
  \bibfield  {author} {\bibinfo {author} {\bibfnamefont {X.}~\bibnamefont
  {{Wu}}}, \bibinfo {author} {\bibfnamefont {P.}~\bibnamefont {{Hu}}}, \bibinfo
  {author} {\bibfnamefont {Z.}~\bibnamefont {{Han}}}, \bibinfo {author}
  {\bibfnamefont {D.~G.}\ \bibnamefont {{Ang}}}, \bibinfo {author}
  {\bibfnamefont {C.}~\bibnamefont {{Meisenhelder}}}, \bibinfo {author}
  {\bibfnamefont {G.}~\bibnamefont {{Gabrielse}}}, \bibinfo {author}
  {\bibfnamefont {J.~M.}\ \bibnamefont {{Doyle}}},\ and\ \bibinfo {author}
  {\bibfnamefont {D.}~\bibnamefont {{DeMille}}},\ }\bibfield  {title} {\bibinfo
  {title} {{Electrostatic focusing of cold and heavy molecules for the ACME
  electron EDM search}},\ }\href@noop {} {\bibinfo  {journal} {in preparation}\
  }\BibitemShut {NoStop}%
\bibitem [{\citenamefont {Masuda}\ \emph {et~al.}(2021)\citenamefont {Masuda},
  \citenamefont {Ang}, \citenamefont {Hutzler}, \citenamefont {Meisenhelder},
  \citenamefont {Sasao}, \citenamefont {Uetake}, \citenamefont {Wu},
  \citenamefont {DeMille}, \citenamefont {Gabrielse}, \citenamefont {Doyle},\
  and\ \citenamefont {Yoshimura}}]{masudasipm}%
  \BibitemOpen
\bibfield  {journal} {  }\bibfield  {author} {\bibinfo {author} {\bibfnamefont
  {T.}~\bibnamefont {Masuda}}, \bibinfo {author} {\bibfnamefont {D.~G.}\
  \bibnamefont {Ang}}, \bibinfo {author} {\bibfnamefont {N.~R.}\ \bibnamefont
  {Hutzler}}, \bibinfo {author} {\bibfnamefont {C.}~\bibnamefont
  {Meisenhelder}}, \bibinfo {author} {\bibfnamefont {N.}~\bibnamefont {Sasao}},
  \bibinfo {author} {\bibfnamefont {S.}~\bibnamefont {Uetake}}, \bibinfo
  {author} {\bibfnamefont {X.}~\bibnamefont {Wu}}, \bibinfo {author}
  {\bibfnamefont {D.}~\bibnamefont {DeMille}}, \bibinfo {author} {\bibfnamefont
  {G.}~\bibnamefont {Gabrielse}}, \bibinfo {author} {\bibfnamefont {J.~M.}\
  \bibnamefont {Doyle}},\ and\ \bibinfo {author} {\bibfnamefont
  {K.}~\bibnamefont {Yoshimura}},\ }\bibfield  {title} {\bibinfo {title}
  {Suppression of the optical crosstalk in a multi-channel silicon
  photomultiplier array},\ }\href {https://doi.org/10.1364/OE.424460}
  {\bibfield  {journal} {\bibinfo  {journal} {Opt. Express}\ }\textbf {\bibinfo
  {volume} {29}},\ \bibinfo {pages} {16914} (\bibinfo {year}
  {2021})}\BibitemShut {NoStop}%
\end{thebibliography}%

\end{document}